\documentclass[useAMS,usenatbib,aps,floatdix,nofootinbib]{mn2e}

\usepackage{color}
\usepackage{graphics}
\usepackage{amssymb,amsmath}
\usepackage{amsmath}
\usepackage{psfrag}
\usepackage{graphicx}

\newcommand{\f}{\frac}

\newcommand{\be}{\begin{equation}}      
\newcommand{\ee}{\end{equation}}      
      
\newcommand{\bef}{\begin{figure}}      
\newcommand{\eef}{\end{figure}}      
\newcommand{\bea}{\begin{eqnarray}}    
\newcommand{\eea}{\end{eqnarray}}

\newcommand{\bx}{{\bf x}}

\def\spose#1{\hbox to 0pt{#1\hss}}
\def\ltapprox{\mathrel{\spose{\lower 3pt\hbox{$\mathchar"218$}}
\raise 2.0pt\hbox{$\mathchar"13C$}}}
\def\gtapprox{\mathrel{\spose{\lower 3pt\hbox{$\mathchar"218$}}
\raise 2.0pt\hbox{$\mathchar"13E$}}}
\def\inapprox{\mathrel{\spose{\lower 3pt\hbox{$\mathchar"218$}}
\raise 2.0pt\hbox{$\mathchar"232$}}}

\def\bx{{\bf x}}  

\def\br{{\bf r}}

\def\bn{{\bf n}}

\def\bse{\begin{subequations}}
\def\ese{\end{subequations}}

\def\bF{{\mathbf F}}

\def\lsim{\raise 0.4ex\hbox{$<$}\kern -0.8em\lower 0.62ex\hbox{$\sim$}} 
\def\gsim{\raise 0.4ex\hbox{$>$}\kern -0.7em\lower 0.62ex\hbox{$\sim$}}

\def\f0N{f_0^{(N)}}
\def\bec{\begin{center}}
\def\eec{\end{center}}

\title[Evolution of isolated overdensities ]
{Evolution of isolated overdensities as a control on cosmological
N body simulations}
\author[M. Joyce and F. Sylos Labini]
{Michael Joyce${^{1}}$ and Francesco Sylos Labini${^{2,}}{^{3}}$\\
$^1$Laboratoire de Physique Nucl\'eaire et de Hautes Energies,
UMR 7585, Universit\'e Pierre et Marie Curie --- Paris 6, \\
75252 Paris Cedex 05, France\\
$^2$Enrico Fermi Center, 
Piazza del Viminale 1
00184 - Rome - Italy\\
$^3$Insitute of Complex Systems (ISC), Consiglio Nazionale delle
Ricerche, Via dei Taurini 19, I-00185 Rome, Italy\\ }
\begin{document}

\date{\today}

\maketitle

\begin{abstract}
Beyond convergence studies and comparison of different codes, there
are essentially no controls on the accuracy in the non-linear regime
of cosmological N body simulations, even in the dissipationless
limit. We propose and explore here a simple test which has not been
previously employed: when cosmological codes are used to simulate an
{\it isolated} overdensity, they should reproduce, in physical
coordinates, those obtained in open boundary conditions without
expansion. {In particular, the desired collisionless nature of the
  simulations can be probed by testing for stability in physical
  coordinates of virialized equilibria}.  We investigate and
illustrate the test using a suite of simulations in an Einstein de
Sitter cosmology from initial conditions which rapidly settle to
virial equilibrium.  We find that the criterion of stable clustering
allows one to determine, for given particle number $N$ in the ``halo"
and force smoothing $\varepsilon$, a maximum red-shift range over
which the collisionless limit may be represented with desired
accuracy. {{We also compare our results to the so-called Layzer Irvine
    test, showing that it provides a weaker, but very useful, tool to
    constrain the choice of numerical parameters.}  Finally we outline
  in some detail how these methods could be employed to test the
  choice of the numerical parameters used in a cosmological
  simulation.}
\end{abstract}

\begin{keywords}
Galaxy: halo; Galaxy: formation; globular clusters: general;
(cosmology:) dark matter; (cosmology:) large-scale structure of
Universe; galaxies: formation
\end{keywords}

\section{introduction}
Numerical simulations of structure formation in the universe in
cosmology use the $N$ body method in which the continuum density field
of dark matter is represented by a finite number of discrete particles
interacting by a smoothed Newtonian two body potential. It is
evidently of importance to control as much as possible for their
precision and reliability. Specifically, beyond issues of numerical
convergence, it is important to understand the limits imposed on the
accuracy of results by the use of a finite number of particles to
represent the theoretical continuum density field, and the associated
introduction of a smoothing scale (or equivalent) in the gravitational
force.  This latter scale, $\varepsilon$, clearly imposes a lower
limit on the spatial resolution, so in order to optimize resolution
the question is {\it how small a value of $\varepsilon$ may be
  employed for a given number of particles and starting redshift}.
This question has been the subject of some controversy, notably
concerning whether values of $\varepsilon$ smaller than the initial
inter-particle distance may be employed (see e.g. \cite{splinter_1998,
  knebe_etal_2000, power_etal_2002, heitmann_etal_2005,
  discreteness3_mjbm, romeo08}).

In this article we discuss one way in which cosmological N body codes
may be tested for their reliability which has not been explored
previously.  The idea is based on the simple observation that, applied
to the simulation of an ``isolated'' overdensity (i.e. a finite system
of size much smaller than that of the periodic simulation box), a
cosmological simulation should be equivalent, in physical coordinates,
to one performed in open boundary conditions without cosmological
expansion.  Indeed the only differences between the two should arise
from possible differences in the force smoothing and finite size
effects, both of which are variables on which the physical results of
a cosmological simulation should not depend. {Even without a direct
  comparison with simulations in open boundary conditions, the desired
  collisionless nature of cosmological simulations can be tested for
  by probing whether an isolated virialized structure, corresponding
  to a collisionless equilibrium, remains {\it stable in physical
    coordinates}.}

By ``isolated" we mean that there is no other mass in the periodic box
other than the structure considered, which itself evolves in a region
of a size small compared to that of the box. The structure is
therefore isolated but for the interaction with its ``copies" included
in the infinite system over which the force is summed. We illustrate
with a set of numerical simulations how this required equivalence of
the evolution in codes with and without expansion can be used to
actually determine whether a given choice of numerical parameters for
cosmological simulations is appropriate.  We focus in particular on
the choice of the smoothing length in the force, and show that the
test allows one to determine a range of appropriate values.

To avoid possible confusion it is probably useful to underline 
the distinction between stable clustering as we study it here, 
and the same term as it is frequently discussed in cosmological 
simulations (see, e.g., \cite{efstathiou_88, smith}):  it can 
be postulated \citep{peebles} that, in the strongly non-linear 
regime, structures evolve {\it as if} they were isolated from
the rest of the mass in the universe. If this ``stable clustering
hypothesis" is valid (to a good approximation, on average) it 
leads, when matched with linear theory, to very specific 
predictions for the nature of the correlations in the non-linear
regime\footnote{Specifically starting from power-law initial conditions
it leads to the prediction of a ``stable clustering hierarchy"
characterized by a power-law correlation function of which
the exponent can be determined  \citep{peebles} }. Here, in 
contrast, we will consider by construction the evolution only 
of a single structure, for which stable clustering {\it must} be 
observed if the simulation is reproducing the desired physical
limit.

The article is organized as follows. In the next section we show
explicitly how cosmological codes used to simulate isolated
overdensities in an expanding universe are related, in an appropriate
limit, to non-expanding codes. We then discuss the particular limit of
virialized halos, for which the stationarity in a non-expanding code
corresponds to stable clustering in physical coordinates in the
expanding code.  In the following section we illustrate and
  investigate the test using a set of simulations in an Einstein de
  Sitter (EdS) universe, which differ only in the smoothing length
  employed. We then also consider a set of simulations in which only
  the box size is varied.  In the subsequent section we compare the
  test with the so-called Layzer Irvine test for the evolution of the
  energy in cosmological simulations.  
  In Sect.~\ref{Nature of discreteness effects and parametric scalings}
  we discuss what can be inferred from our results about the role
  of different possible discreteness effects in producing the
  observed deviations from stable clustering, and what can
  be concluded about dependences of these deviations 
  on the relevant parameters (particle number, force smoothing, box size).
  In the following section we specify in  a ``recipe" form how the test could 
  be employed practically by simulators to test choice of numerical parameters 
  in cosmological simulations.  We conclude with a brief discussion
  of possible variants on the tests and some more general comments.
 
\section{Evolution of an isolated overdensity in cosmological N body codes}

Dissipationless cosmological N-body simulations (see, e.g.,
\cite{kravtsov_1997, springel_etal_2001, teyssier_2002}, or
\cite{bagla_review} for a review) solve numerically the equations
\begin{equation}
\frac{d^2 {\bf x}_i}{dt^2} +
2H \frac{d{\bf x}_i}{dt}  = \frac{1}{a^3} \bF_i 
\label{3d-equations-1}
\end{equation}
where
\begin{equation}
\bF_i  =- Gm \sum_{j \neq i}^P
\frac{{\bf x}_i - {\bf x}_j}{\vert {\bf x}_i - {\bf x}_j \vert^3} 
W_\varepsilon (\vert {\bf x}_i - {\bf x}_j \vert) \,.
\label{3d-equations-2}
\end{equation}
${\bf x}_i$ are the comoving particle coordinates of the $i=1...N$
particles of equal mass $m$, enclosed in a cubic box of side $L$, and
subject to {\it periodic boundary conditions}, $a(t)$ is the
appropriate scale factor for the cosmology considered, and $H(t)={\dot
  a}/{a}$ is the Hubble constant. The function $W_\varepsilon$ is a
regularisation of the divergence of the force at zero separation ---
below a characteristic scale, $\varepsilon$, which is typically fixed
in comoving units. For simplicity we will drop this function {\it in
  this section} as it plays no role for our considerations here.

The superscript `P' in the sum in (\ref{3d-equations-2}) indicates
that it runs over the infinite periodic system, i.e., the force on a
particle is that due to the $N-1$ other particles {\it and all their
  copies}. The sum, as written, is formally divergent, but it is
implicitly regularized by the subtraction of the contribution of the
mean mass density.  This is physically appropriate, in an expanding
universe, as the mean mass density sources the expansion (see
e.g. \cite{peebles}).

\subsection{Equations of motion a single ``isolated" structure in physical coordinates}

\begin{figure}
\vspace{1cm}
{
\par\centering \resizebox*{9cm}{8cm}{\includegraphics*{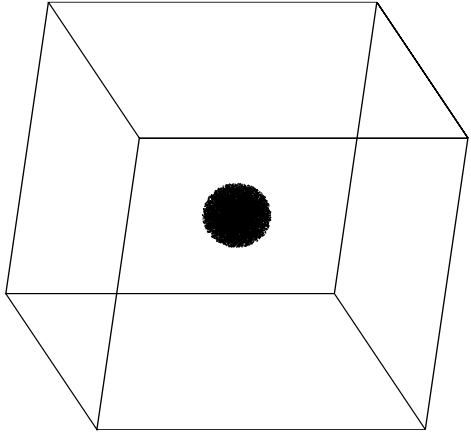}}
\par\centering
}
\caption{Schematic representation of the case studied in this paper:
a single structure of $N$ particles confined in a region of characteristic
size small compared to the side of the cubic cosmological simulation 
box (dark line).} 
\label{fig1} 
\end{figure}

Let us now consider the case illustrated schematically in Fig.\ref{fig1}  
where the $N$ particles are contained within a
spherical volume, $\Omega$, of radius $R$, with $R<L/4$ 
(where $L$ is the side of the cube).
The latter condition is sufficient to ensure that the
distance between any particle $i$ and any other particle $j$ in
$\Omega$ is less than that separating $i$ from any copy of $j$
in the infinite periodic system. The force on a particle $i$
may then be written

\begin{equation}
\bF_i=  \bF_i^{\Omega} + \bF_i^{c}
\end{equation}
where 
\begin{equation}
 \bF_i^{\Omega}= -Gm \sum_{j \neq i} \frac{{\bf x}_i - {\bf
     x}_j}{\vert {\bf x}_i - {\bf x}_j \vert^3}
\end{equation}
is simply the direct one over the $N-1$ other particles in the volume
$\Omega$, and
\begin{equation}
\label{eq5}
\bF_i^{c}=  \sum_{j \neq i} \bF_{ij}^c= - Gm \sum_{j \neq i}  \sum_{\bn \neq 0} 
\frac{{\bf x}_i - {\bf x}_j - \bn L}{\vert {\bf x}_i - {\bf x}_j - \bn L \vert^3}
\end{equation}
is the force due to all the particles in the copies, labelled by all
vectors of non-zero integers $\bn$, and where we have $\vert {\bf x}_i
- {\bf x}_j \vert < L/2 < \vert \bn L \vert$.  Note that
$\bF_i^{\Omega}$ is clearly finite and well defined, while each sum
over $\bn$ giving $\bF_{ij}^c$ is now formally divergent, but again
implicitly regulated by the subtraction of the mean density.

To calculate $\bF_{ij}^c$ we observe that it corresponds simply to the
force on a {\it single} particle displaced by $(\bx_i-\bx_j)$ off an
infinite perfect lattice of lattice spacing $L$. It is straightforward
to show (see \cite{gabrielli_06}) that, expanding in Taylor series in
$(\bx_i-\bx_j)$ about ${\bf x}_i - {\bf x}_j=0$, we have
\begin{equation}
\bF_{ij}^c= \frac{4\pi G \rho_0}{3N} (\bx_i - \bx_j) + O 
(  \vert \bx_i - \bx_j \vert^2 /L^4)
\label{pert-lattice-force}
 \end{equation}
where $\rho_0=mN/L^3$ is the total mean mass density (and thus
$\rho_0/N$ the mean mass density of the lattice of the particle $j$
and its copies). The leading non-zero ``dipole'' term on the right hand side in 
(\ref{pert-lattice-force}) is a repulsive term which arises from
the subtraction of the mean mass density to
regulate the sum: it is precisely the force arising from the
mass contained in a sphere of radius $\vert {\bf x}_i - {\bf x}_j \vert$
of constant mass density $-\rho_0/N$. We do not write explicitly
the sums for the subsequent (quadrupole and higher
multipole) terms, but they are manifestly convergent and
suppressed by positive powers of $(R/L)$ compared to the 
dipole term.

As the sum over $\bx_j$ in $\bF_i^{c}$ vanishes because we have chosen
(without loss of generality) to place the center of mass of the $N$
particles in $\Omega$ at the origin of our coordinates, retaining only
the dipole term in $\bF_i^{c}$ the equations of motion
(\ref{3d-equations-1}) we obtain
\begin{equation}
\frac{d^2 {\bf x}_i}{dt^2} +
2H \frac{d{\bf x}_i}{dt}  = -\frac{Gm}{a^3}  \sum_{j \neq i}  \frac{{\bf x}_i - {\bf x}_j}{\vert {\bf x}_i - {\bf x}_j \vert^3} 
 + \frac{4\pi G \rho_0}{3} \bx_i  
\label{3d-equations-3}
\end{equation}
where the sum is now only over the $N-1$ particles in $\Omega$.
Assuming an EdS cosmology, for which 
\begin{equation}
\frac{\ddot{a}}{a} = - \frac{4\pi G \rho_0}{3 a^3}\,,
\end{equation}
these equations (\ref{3d-equations-2}) may be written, in physical
coordinates $\br_i \equiv a(t) \bx_i$, simply as
\begin{equation}
\frac{d^2 {\bf r}_i}{dt^2} = -Gm \sum_{j \neq i} \frac{{\bf r}_i -
  {\bf r}_j}{\vert {\bf r}_i - {\bf r}_j \vert^3}
\label{3d-equations-4}
\end{equation}
i.e. as the equations of motion of $N$ isolated purely
self-gravitating particles~\footnote{In the case of a cosmology with
  matter and a cosmological constant $\Lambda$, we obtain in physical
  coordinates an additional repulsive term arising from
  $\Lambda$. This can easily be incorporated in the considerations
  which follow, but we consider here for simplicity the case
  $\Lambda=0$, i.e., the EdS cosmology.}.

Thus, when a cosmological code is used to simulate an isolated
structure in an expanding universe, it should reproduce the same
result, in physical coordinates, as that obtained for such a structure
in open boundary conditions without expansion. This identity is valid
up to
\begin{itemize}
\item finite size corrections, arising from the use of a finite
  (periodic) box in cosmological simulations (and which vanish in the
  limit $R/L \rightarrow \infty$).
\item eventual differences due to force softening in the two type of
  codes (which we have neglected above).
\end{itemize}
 
Any dependence of results on the box size or force smoothing is
unphysical in cosmological codes. Thus these codes can be tested by
using them to simulate isolated structures and comparing the results
obtained to those for the same initial conditions in open boundary
conditions and without expansion.

\subsection{Virialization and stable clustering}

Results for the detailed evolution from arbitrary initial conditions
in open boundary conditions and without expansion can be obtained in
general only from numerical simulation. However, even without
performing such simulations, it is possible to do tests of
cosmological codes which are based on well established generic
features of the evolution in open boundary conditions.  One such
feature, for a very wide class of simple initial conditions, is the
evolution to a virial equilibrium in a few dynamical times (see
e.g. \cite{heggie+hut_book}).  These equilibria are, in the limit $N
\rightarrow \infty$, stationary states corresponding to time
independent solutions of the collisionless Boltzmann equation. In a
finite $N$ system they evolve away from this equilibrium, on a very
long time scale diverging with $N$, due to collisional effects.  {The
  stationarity of these states in the collisionless limit corresponds,
  in a cosmological code, to ``stable clustering'' of the virialized
  system.}

It follows that by studying the stability of the evolution obtained
from appropriate initial conditions, we can test cosmological codes
both for effects arising from the finite size of the box, and force
smoothing, as well as for collisional effects.  These are precisely
the principle undesired effects introduced by using the N-body method
to solve the (continuum, infinite system) cosmological problem of
formation of structure, and present even in the idealized limit that
the numerical integration of the equations of motion is arbitrarily
accurate.  Thus by testing for the validity of stable clustering in
such a regime we can test for the capacity of N body codes to
reproduce correctly the relevant physical limit. It is such a test
which we will focus on in what follows.

\section{Numerical study of the test}
\label{TestCase}

To illustrate the test we use the GADGET code ~\footnote{See {\tt
    http://www.mpa-garching.mpg.de/gadget/} } \citep{springel_05},
which can be used to perform both cosmological simulations, and
simulations without expansion (in both open and periodic boundary
conditions). For our cosmological simulations we consider, for
simplicity, evolution in an Einstein de Sitter background. {All our
  simulations here are for $N=10^4$ particles}.

\begin{figure}
\vspace{1cm}
{
\par\centering \resizebox*{9cm}{8cm}{\includegraphics*{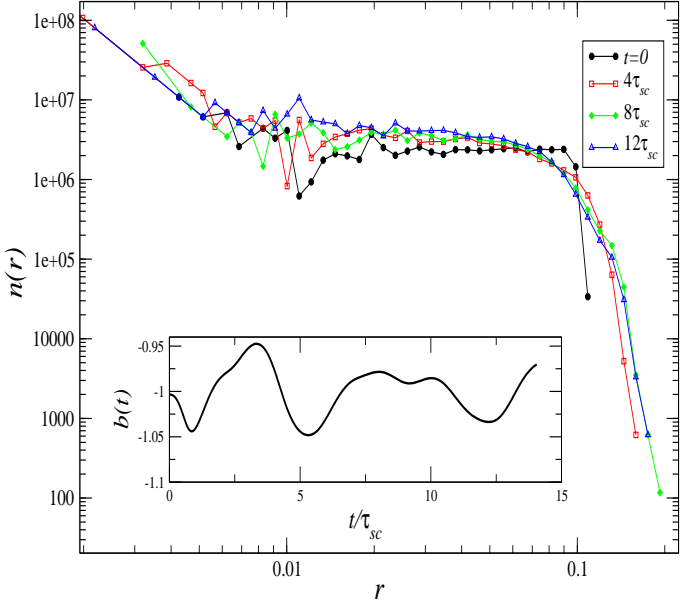}}
\par\centering
}
\caption{Density profile (and, inset, virial ratio) at indicated times
for our chosen initial conditions when evolved in open boundary 
conditions (and without expansion).} 
\label{fig2} 
\end{figure}

\subsection{Initial conditions and choice of units}
\label{Initial conditions and choice of units}

The initial conditions we study here are the following: the $N=10^4$ 
particles are randomly  distributed in a spherical volume of radius 
$R=0.1L$, and assigned random velocities sampled
from a probability distribution which is {\it uniform} in a cube 
centered at zero velocity. {These velocities are then normalized so that 
$b=-1$, where 
\begin{equation}
b=\frac{2K_p}{U_p}
\label{virialization}
\end{equation}
is the virial ratio, and $K_p$ and $U_p$ are the peculiar kinetic
energy and peculiar potential energy respectively. These
are defined by 
\be
K_p = \frac{1}{2}  \sum_i m |\vec{v}_i|^2
\label{peculiar-K}
\ee
where $\vec{v_i}= a(t) \frac{d \vec{x}_i}{dt}$ is the 
particle peculiar velocity, i.e., 
\be
\vec{v}_i= a(t) \frac{d \vec{x}_i}{dt}= \frac{d \vec{r}_i}{dt} - H(t) \vec{r}_i
\label{def-PecVel}
\ee
and 
\be
U_p = \frac{1}{2a} \sum_{i,j, i\neq j}  m g (|\vec{x_i} -  \vec{x_j}|)
\label{peculiar-U}
\ee where $g(r)$ is the exact (GADGET) two body potential. Thus,
modulo force smoothing, $U_p$ is equal to the Newtonian potential
energy in physical coordinates, and we will therefore refer to it as
the {\it physical potential energy}
\footnote{{Note that $g(r)$ differs 
from the exact two body potential used in the
dynamical evolution, because of the modifications
associated with the periodic  boundary conditions; 
often the nomination ``peculiar  potential energy"  is 
used for $U_p$ defined as in Eq.~(\ref{peculiar-U}) 
but including this modification in $g(r)$.}}. }

Our motivation for this choice is that it is a simple one which,
although out of equilibrium, rapidly settles to a virial
equilibrium. We note that it corresponds, in the cosmological
simulations, to an initial density fluctuation of amplitude
\begin{equation}
\delta = \frac{\rho}{\rho_0} = \frac{3L^3}{4\pi R^3} \approx 240
\label{ratio-densities}
\end{equation}
where $\rho_0$ is the mean (comoving) mass density of the
universe.Thus it can be thought of, roughly, as representing an almost
virialized spherical halo at its formation time, which is then evolved
forward in isolation from the rest of the mass in the
universe \footnote{{The virial condition $b=-1$ as given indeed
    corresponds, to a very good approximation, to the more evident
    condition in physical coordinates: $2K_p=-U_p$ implies ${v}_i^2
    \sim GM/R$, from which it follows that ${v}_i^2/H^2 r_i^2 \sim
    \delta$, i.e., the peculiar velocity is large compared to the
    Hubble flow velocity. }}.  In our conclusions we will briefly
discuss other initial conditions which it would be relevant to study
in testing cosmological simulations.

The results we report require only choice of units for length 
and energy: for the former we will take units defined by $L=1$, 
and for the latter units in which $(3GM^2/5R)=1$, i.e. in 
which the initial continuum limit potential energy is (minus) 
unity.  

We note that (\ref{ratio-densities})  implies that, at 
expansion factor $a$, starting from $a=1$ at 
$t=t_0=1/\sqrt{6 \pi G \rho_0}$,  we have 
\begin{equation}
t-t_0=[a^{3/2}-1] t_0 = \frac{4}{3\pi} \sqrt{\delta} \, [a^{3/2} -1]
\tau_{sc}\approx 6.6 \, [a^{3/2} -1] \tau_{sc}
\label{time-elapsed}
\end{equation}
where $\tau_{sc}=\sqrt{3 \pi/ 32 G \rho}$ is the collapse time
for a cold uniform sphere with mass density $\rho$.

Our expanding simulations are evolved up to a scale 
factor $a=20$,  unless otherwise indicated. This 
means our study is (roughly) of halos of $N=10^4$
particles which formed at a red-shift less than about twenty. Note that
(\ref{time-elapsed}) implies that $a=20$ corresponds
to several hundred dynamical times of the halo.
For $N=10^4$ particles this is sufficiently long,
as we will see, to see evidence large deviations 
from stable clustering in all our simulations. 

\subsection{Parameters of expanding simulations}
\label{Parameters of expanding simulations}

We consider a set of simulations (of $N=10^4$ particles) with
the values of smoothing  $\varepsilon$ shown in Table~\ref{table-one}. 
In GADGET the smoothed two body potential has a complicated functional
form which is a spline interpolation between the exact Newtonian
potential --- above a separation of $2.8 \varepsilon$ --- and a
potential of which the first derivative vanishes at zero separation. The
value of $\varepsilon$ we quote is the value of the parameter with
this name in the code. At this separation the smoothed potential is
approximately $75 \%$ of it Newtonian value, while at $\varepsilon/2$
it is down to approximately $50 \%$.

{The values of the  parameters controlling precision of the time stepping 
and force calculations are given in Appendix \ref{appendix}, as well as 
a discussion of tests we have performed for the sensitivity of our results 
to variation of these parameters.

Each row in Table~\ref{table-one} gives the name of a simulation and
the corresponding value of $\varepsilon$ in units of the box size $L$.
Also given is the ratio of $\varepsilon$ to $\ell=L N^{-1/3}$.  The
latter corresponds to the initial grid spacing of a cosmological
simulation with the same mean (comoving) matter density.  The {\it
  initial mean nearest neighbour separation} $\Lambda$, on the other
hand, is given by}~\footnote{This expression is derived from the value
  for an infinite Poisson distribution (for which the exact nearest
  neighbour distribution is known, see e.g. \cite{book}).}  
\be
\Lambda = 0.55 \times \left(\frac{4 \pi R^3}{3 N}\right)^{1/3} \approx
0.9\, \ell \, \frac{R}{L} \,
\label{definition}
\ee
{i.e. $\varepsilon/\Lambda \approx 9 \varepsilon/ \ell$  since $R=0.1 L$.}

\begin{table}
\begin{center}
\begin{tabular}{|c|c|c|c|c|c|}
\hline
  Name & $\varepsilon/L$  & $\varepsilon/ \ell$ \\
\hline 
  s1           & 3.7 $\times$ 10$^{-5}$     & 0.0008    \\
  s2         & 5.0 $\times$ 10$^{-5}$     & 0.001 \\
  s3          & 8.0 $\times$ 10$^{-5}$     & 0.0016 \\
  s4  & 1.0 $\times$ 10$^{-4}$     & 0.002  \\
  s5           & 3.7 $\times$ 10$^{-4}$     & 0.008  \\
  s6          & 5.0 $\times$ 10$^{-4}$     & 0.01   \\
  s7           & 3.7 $\times$ 10$^{-3}$     & 0.08  \\
\hline
\end{tabular}
\end{center}
\caption{ Names and corresponding values of $\varepsilon$  of simulations
with $N=10^4$ particles. $L$ is the box size and  $\ell=L N^{-1/3}$. 
}
\label{table-one}
\end{table}

{  We thus consider in all our simulations, as in many
large cosmological simulations, a smoothing which is fixed in comoving units. 
As we will discuss a little more in our conclusions, our test can of course be 
applied with any  other smoothing prescription, e.g., fixed smoothing
in physical coordinates or adaptative smoothing. } 
{The motivation for the range of $\varepsilon$ we have chosen to
study, which extends to values significantly smaller to
those typically used in large cosmological simulations, is the following:

\begin{itemize}
\item An {\it upper} cut-off on $\varepsilon$ is imposed
by the fact that this scale must be sufficiently small so that
gravitational mean field forces may be well approximated.
This requires simply that 
\begin{equation}
\varepsilon \ll R_s\,
\label{epsilon-upper-bound}
\end{equation}
where $R_s$ is the characteristic scale on which the mass density in
the structure varies (in the continuum limit).  The value of
$\varepsilon$ in s7 corresponds to $\varepsilon \approx 0.04 \,R$.
Given that $\varepsilon$ is fixed in comoving coordinates, while
stable clustering will lead to a structure with $R_s \propto 1/a$,
this simulation (run up to $a=20$) will clearly be expected to
manifest the effects of the violation of the bound
(\ref{epsilon-upper-bound}).

{ We note that even the largest value of $\varepsilon$ we consider
  corresponds to a value significantly smaller than $\ell$. Indeed
  such a choice is unavoidable if one wishes to resolve the non-linear
  evolution of structures with a modest number of particles in a
  cosmological simulation.  As mentioned in our introduction, the use
  of $\varepsilon < \ell$ has been the source of discussion and
  controversy in the literature, notably as at early times it leads
  inevitably to effects which should be absent in the desired
  mean-field limit (see e.g. \cite{splinter_1998, discreteness3_mjbm,
    romeo08}).  Our present test, which considers only the strongly
  non-linear regime subsequent to collapse and virialization, clearly
  cannot give us any information or constraint on the accuracy with
  which collisionless behaviour is reproduced in the early time
  evolution. We will return briefly to these issues in our
  conclusions.}

\item
A {\it lower} cut-off on the other hand is dictated in principle only
by numerical limitations: without such a cut-off hard two body
collisions will occur, with (arbitarily) large accelerations requiring
integration with correspondingly small time steps (for a discussion,
see e.g. \cite{knebe_etal_2000}). Given that a two body collision is
soft if $ (Gm/sv_r^2) \ll 1$, where $v_r$ is the initial relative
velocity and $s$ the impact factor, a {\it naive} estimate of the
condition on softening to suppress hard collisions is $\varepsilon \gg
(Gm/v^2)$, where $v$ is the typical speed of a particle in the
system. For a virialized system of $N$ particles of size $R_s$ we have
$Nv^2 \sim GmN^2/R_s$. Thus we estimate
\begin{equation}
\varepsilon  \gg R_s/N \,.
\label{epsilon-lower-bound-1}
\end{equation}
The value of $\varepsilon$ in s1 corresponds 
approximately to this estimated lower bound at the  
beginning of the simulation. Note that, in contrast to the
upper bound   Eq.~(\ref{epsilon-upper-bound}), an evolution 
corresponding to stable clustering (with $R_s \sim 1/a$) 
will improve progressively the satisfaction of
the condition Eq.~(\ref{epsilon-lower-bound-1}).

\end{itemize}

Given that the goal of N body simulations is to reproduce the
collisionless limit, in which hard two body collisions should clearly
play no role, the imposition of the lower bound
(\ref{epsilon-lower-bound-1}) is clearly justified.  Indeed we note
that simple estimates of the minimal $\varepsilon$ required to
reproduce the collisionless limit suggest that much larger values of
$\varepsilon$ may be necessary.  If one assumes, for example, that
$\varepsilon$ should be large enough so that the force from a single
particle is always subdominant with respect to the mean field force,
one obtains
\begin{equation}
\varepsilon \gg R_s N^{-1/2} \,.
\label{epsilon-lower-bound-2}
\end{equation}
An even more restrictive condition, that might possibly
be relevant, is that $\varepsilon$ should be at least
of order the average interparticle distance, i.e., 
\begin{equation}
\varepsilon \gg R_s N^{-1/3} \,.
\label{epsilon-lower-bound-3}
\end{equation}
As we will discuss further below, although we do not do so here
the test we develop could in principle be used to determine 
which (if any) of these scalings is the right one for the 
minimal $\varepsilon$.

\subsection{Evolution in open boundary conditions}
\label{Evolution in open boundary conditions}

As discussed above, the test we explore here, for stability in
physical coordinates of a virialized structure, does not necessarily
require direct comparison with the same initial conditions evolved in
open boundary conditions. Such comparison constitutes, however, a more
stringent test, and, as we will see, can be used to derive more
precise quantitative conclusions about the choice of numerical
parameters.

We have therefore evolved the initial conditions described above in
open boundary conditions without expansion \footnote{See, e.g.,
  \cite{tirawut-thesis, sl2012} for a more detailed study of this
  case.}. The values of the numerical parameters we have used, and
tests we have performed for the stability of these results, are given
in Appendix \ref{appendix}.  Shown in Fig.~\ref{fig2} is the initial
and evolved density profile obtained at the indicated times in a
simulation in open boundary conditions and without expansion. In inset
is shown the evolution of the virial ratio $b$ (defined as
Eq.~(\ref{virialization}), taking $a(t)=1$).  After a few dynamical
times the system has settled down ``gently" to a macroscopically
stable virialized configuration, with small residual fluctuations
about it. Below we will use a smooth fit to the profile at $12
\tau_{sc}$ as a template for the profile of the {\it collisionless}
equilibrium established in open boundary conditions from these initial
conditions. We will then compare this template with the profile of the
virial equilibrium obtained in the cosmological code.  In absence of
collisional effects, finite size effects or other numerical effects,
the two should coincide. Note that one could also perform a test in
which the open system is evolved to the considerably longer times on
which collisional effects play a role, and compare this with the
evolution in the expanding case. This, however, is not a test we
explore here as our focus is on testing the collisionless nature of
cosmological simulations.}

\subsection{Evolution in cosmological code}
\label{Cosmological codes: variation of force smoothing}

\subsubsection{Virial ratio}
\begin{figure}
\vspace{1cm}
{
\par\centering \resizebox*{9cm}{8cm}{\includegraphics*{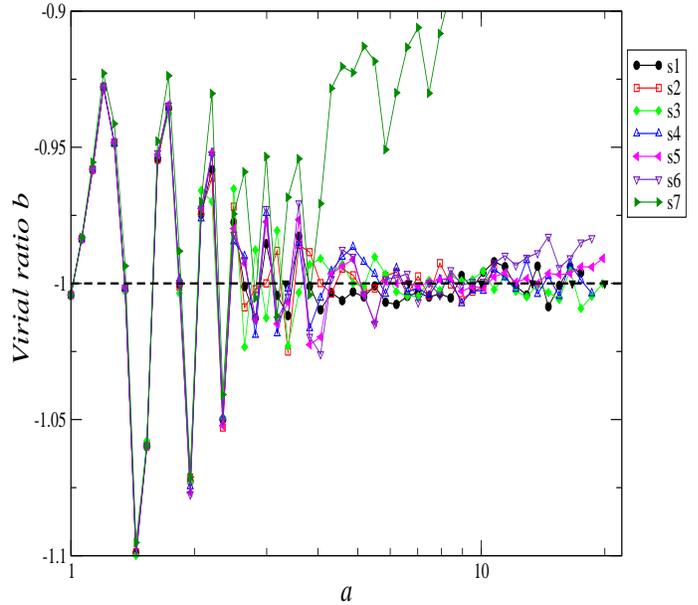}}
\par\centering
}
\caption{Evolution of the virial ratio $b$ (as defined in the text) up
  to scale factor $a=20$, for the different simulations in
  Table~\ref{table-one}.}
\label{fig3}
\end{figure}

In Fig.~\ref{fig3} are shown for the indicated simulations, the
evolution, as a function of redshift, of the virial ratio, as defined
by Eqs.~(\ref{virialization}), (\ref{peculiar-K}) and
(\ref{peculiar-U}).

For all the simulations, except s7, the virial ratio evolves
qualitatively {as would be expected if the system evolves as in open
  boundary conditions: they show} low amplitude coherent oscillations
which decay gradually indicating virialisation (corresponding to
$b=-1$).  Further we have checked that there is good quantitative
coherence between the amplitude and time scale of these oscillations,
and those found in open boundary conditions (inset of
Fig.~\ref{fig2}): using Eq.~(\ref{time-elapsed}) we see that the
temporal range of the latter corresponds just to evolution up to $a
\approx 2$ in the expanding case. We will consider below the exact
degree of agreement between the density profiles obtained in the
expanding simulations and the non-expanding case.

The fact that s7 behaves so differently --- deviating clearly from a
behavior like that in the other cases at $a \approx 5$ --- can be
attributed, { as anticipated above, to the violation of the constraint
  (\ref{epsilon-upper-bound}): } we have $\varepsilon/R_s \approx
0.04$ initially, which means that, if the structure does indeed remain
fixed in physical coordinates, at $a \sim 5$ the effective {\it
  mean-field} force due to all particles is very different to its
Newtonian value. For s6, on the other hand, with a smoothing about
seven times smaller than in s7, we expect to have $\varepsilon/R_s
\approx 0.1$ at $a=20$, still small enough so that the Newtonian value
of the mean field potential is well approximated by the smoothed
potential. In this latter case, however, even from $a \approx 10$ the
virial ratio already appears in Figs.\ref{fig2} and \ref{fig3} to show
a slight tendency to increase at the larger values of $a$.

\subsubsection{Potential energy}

\begin{figure}
\vspace{1cm} { \par\centering
  \resizebox*{9cm}{8cm}{\includegraphics*{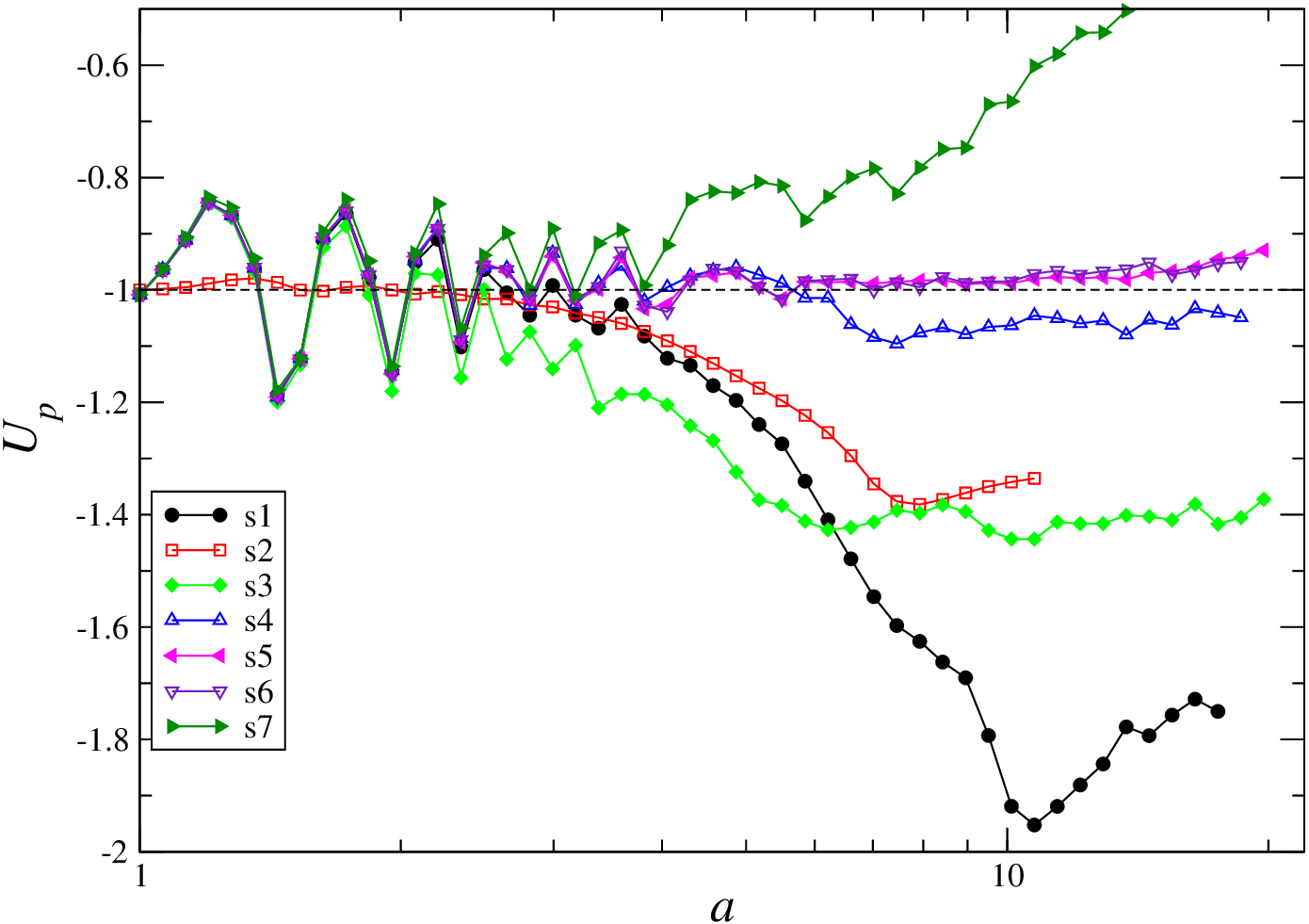}}
  \par\centering }
\caption{Evolution of the potential energy $U_p$ of the
  structure (in physical coordinates), for the indicated simulations
  in Table~\ref{table-one}.}
\label{fig4}
\end{figure}

In Fig.~\ref{fig4} are shown the evolution of the 
physical potential energy $U_p$. If the structure, once virialized, remains
stable in physical coordinates we should have $U_p= constant$.
Further, of course, {\it if the smoothing plays no role, this constant
  value should be the same in all the simulations}. We observe that
only the behavior observed for the simulations s5 and s6 appears
to be consistent with stable clustering of the virialized structure,
and even in these two cases a slight deviation is apparent at the end
of simulated red-shift range, from $a \approx 18$. These plots thus
indicate that at most in the corresponding narrow range of
$\varepsilon$ can the behavior required in the desired continuum
limit be reproduced by the $N$ body method.

\subsubsection{Density profiles}

\begin{figure*}
\vspace{1cm} { \par\centering
 \resizebox*{18cm}{16cm}{\includegraphics*{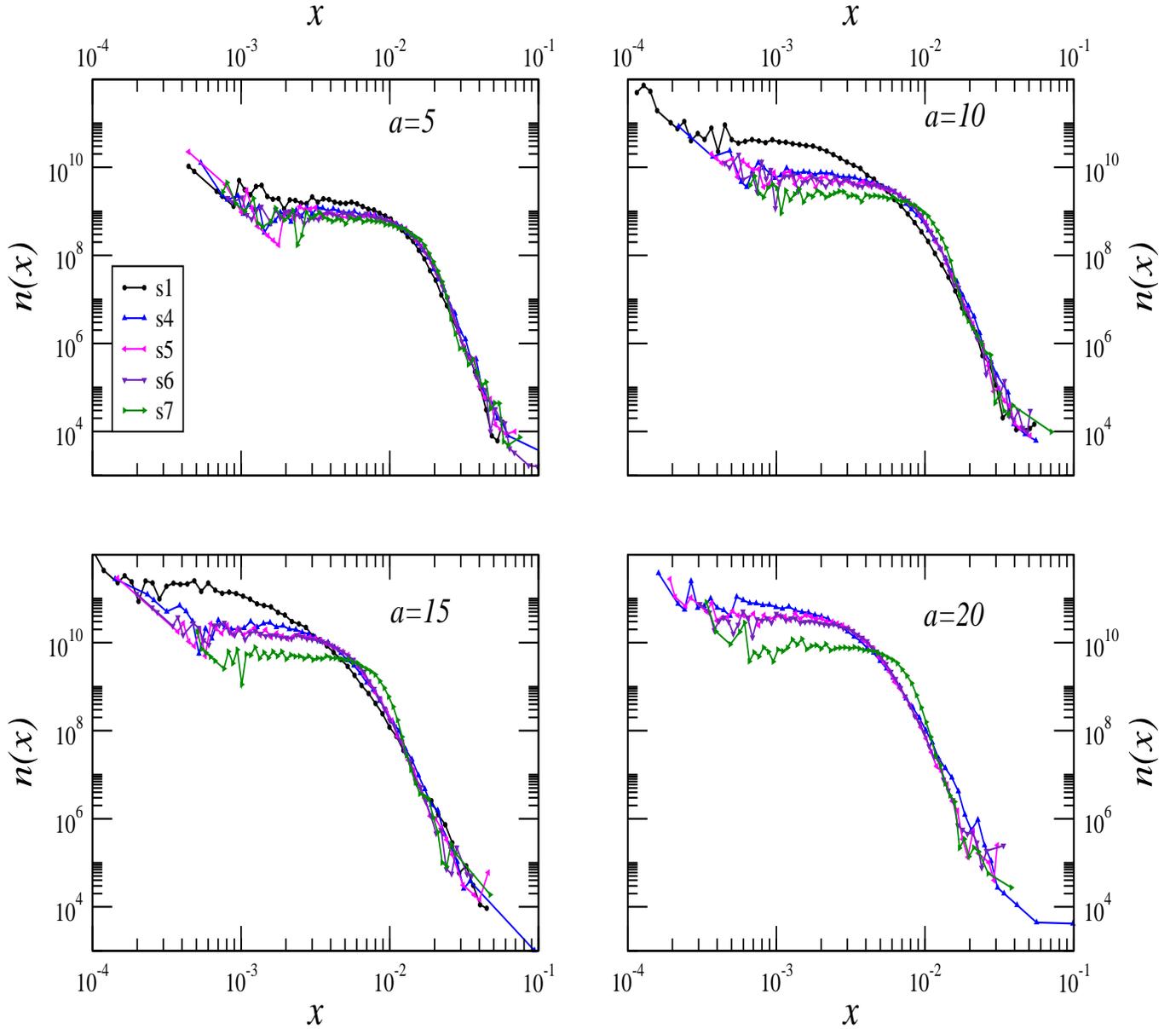}}
  \par\centering }
\caption{Density profiles in comoving coordinates as measured 
in the indicated simulations from Table~\ref{table-one}, at the
different indicated scale factors.}
\label{fig5}
\end{figure*}

Let us now examine whether this conclusion is borne out by further
analysis of the evolved configurations.  Shown in Fig.~\ref{fig5} are
the measured density profiles at the indicated scale factors in each
of the simulations. The results confirm strongly what can be
anticipated from the analysis of the potential energy: the profiles
agree increasingly poorly in time, with s7 and s1 clearly giving
profiles completely different to those obtained in the other cases.
s4, on the other hand, shows a much smaller discrepancy with the
remaining two, s5 and s6. These latter two simulations agree very well
with one another, except at the very last plot where a slight
difference may be observed. We note that, compared to these cases, the
simulations with a smaller $\varepsilon$ have a denser more
concentrated core, while for s7, which has a larger value of
$\varepsilon$, the opposite behavior is observed.  This is very
consistent with our comments above about this latter simulation:
$\varepsilon$ is so large that mean field forces are very reduced
compared to the exact Newtonian mean field, leading to a much less
condensed structure.


\subsubsection{Rescaled density profiles}

\begin{figure*}
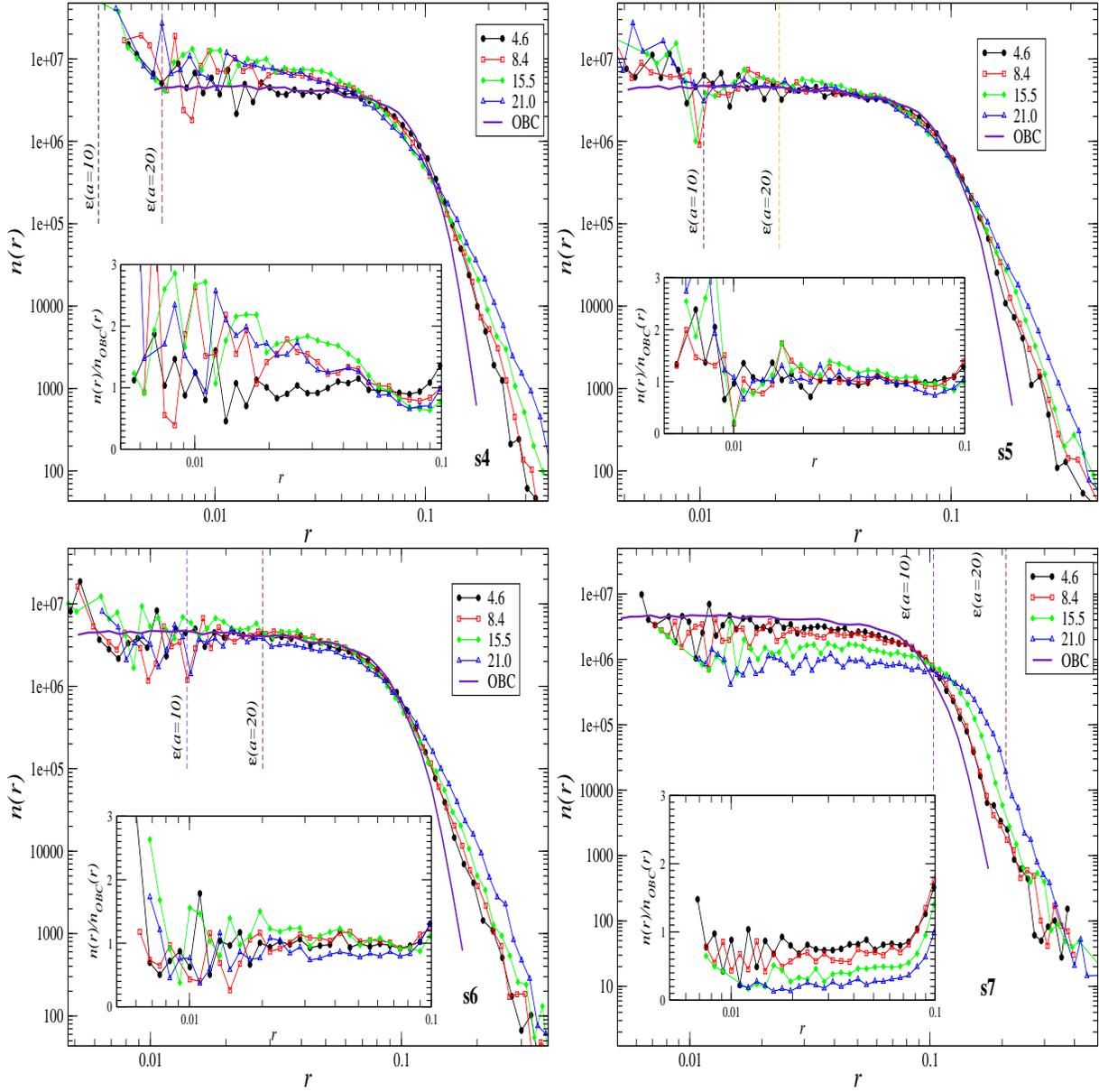

\vspace{1cm}
{
\par\centering
\resizebox*{8cm}{8cm}{\includegraphics*{Fig6a.eps}}
\resizebox*{8cm}{8cm}{\includegraphics*{Fig6b.eps}}
\resizebox*{8cm}{8cm}{\includegraphics*{Fig6c.eps}}
\resizebox*{8cm}{8cm}{\includegraphics*{Fig6d.eps}}
\par\centering
}
\caption{Density profiles in physical coordinates 
at the different indicated values of $a$. Each figure corresponds
to the single indicated simulation. Also shown in each case
is a smooth fit to the profile (labelled OBC) obtained in an 
open non-expanding simulation at $t=12\tau_{sc}$. In the inset
plots of the profiles normalized by this latter profile are given. }
\label{fig-rescaled-profiles}
\end{figure*}

It can be seen qualitatively from the previous figures (which are
plotted in comoving coordinates) that the comoving size of the
structures does indeed decrease, with a corresponding increase in
their density.  To see whether the behavior is quantitatively in
agreement with that associated with stable clustering, we show in
Fig.~\ref{fig-rescaled-profiles} the evolution of the profile in
physical coordinates for the simulations indicated, i.e., we plot in
each case $n(r)=n(x)/a^3$ as a function of $r=a x$. In this
representation stable clustering corresponds to an invariant
profile. {We also show in these plots the template for the profile of
  the collisionless equilibrium obtained from a simulation of the open
  non-expanding case, as described in Sect.~\ref{Evolution in open
    boundary conditions}. } The insets in the plots show the results
in each expanding simulation normalized to this profile.

The results confirm what has been anticipated above from the
examination of the behaviour of $U_p$ and the comparison of density
profiles, but also give additional constraints: s5 and s6 reproduce
stable clustering considerably better than any of the other
simulations, but s5 also clearly does better than s6, which shows a
deviation between $a=15$ and $a=20$. Thus, of the full range of
$\varepsilon$ considered, s5 is closest to optimal, while all others
lead to quite measurable deviations from the continuum limit
behaviour. {We see very clearly in these plots again the marked
  qualitative difference between the cases of a smoothing which is too
  large and one which is too small.  In the former case the structure
  obtained is very much less dense and more extended than it should
  be, while in the latter case it is very much more compact, with a
  density profile which steepens towards the centre.  These
  differences, as we will discuss further in the final section below,
  clearly correspond to the very different effects at play in the two
  cases.}

We note that the conclusions in the previous paragraph can be drawn
even without the direct comparison with the non-expanding density
profile: in other words, when stability is observed in a given
expanding universe simulation, the (stable) profile obtained is always
consistent with the correct one. The comparison with the non-expanding
profile, shown in the inset of each figure gives a more quantitative
measure of the deviation of the corresponding expanding universe
simulation from the correct behavior. Note that these insets have been
cut in all cases at $r=0.1$, since {\it in all cases} there are very
significant deviations beyond this radius, which is reflected also
clearly in clear deviations from stable clustering.  {Indeed we note
  that in all cases the very outer part of the profile extends very
  significantly further than in open boundary conditions. In most
  cases this corresponds to a very small fraction of the total mass,
  except in the case of s7 (with the largest smoothing) for which the
  characteristic size of the whole structure is, as we have noted,
  very much larger than it should be.}

\subsection{Test for box size dependence}
\label{Results for different box sizes}

{The simulations in the previous section are of fixed box size. The
  fact that they evolve differently is, by construction, due only to
  the different force smoothing. However when we compare the profiles
  obtained to the template in open boundary conditions (determined at
  very early times and taken to be representative of collisionless
  evolution), differences may also arise because of the periodic
  boundary conditions. As we have explained the differences with
  respect to the open system should be suppressed by powers of $R/L$
  and thus vanish as the size of the system becomes small compared to
  the box size. By varying the box size we can test both whether
  finite size effects may be observed, and whether they are, as we
  have assumed, small for $R/L=0.1$ }

{To do so we have run the set of simulations in
  Table~\ref{table-three}.  The simulation s5R0.1 is identical to the
  simulation s5 considered above, except that it has been run for a
  longer time, up to $a=33$ (rather than $a=20$). The other four
  simulations are for the same number of particles, $N=10^4$, and
  differ only in their initial size $R$.  Because the overdensity
  $\delta$ represented by these initial conditions depends on $R$
  [cf.Eq.~(\ref{ratio-densities})], the relation
  Eq.~(\ref{time-elapsed}) between the physical time elapsed and the
  scale factor $a$ is modified. More precisely, in units of the
  characteristic time of the isolated structure $\tau_{sc}$, a given
  elapsed time $t-t_0$ corresponds to a fixed value of $R^{-3/2} \,
  (a^{3/2} -1)$. The scale factor $a=a_{\rm end}$ in
  Table~\ref{table-three}, up to which the corresponding simulation
  has been run, has been chosen so that $R^{-3/2} \, (a_{\rm
    end}^{3/2} -1)$ is equal in all cases, i.e., all simulations are
  run up to the same time in units of $\tau_{sc}$.

The choice of $\varepsilon$ given in Table~\ref{table-three}
have been made by scaling the value in s5 in proportion to 
$R$ (and the mean interparticle distance). As $\varepsilon$
is fixed in comoving coordinates, it evolves in physical
coordinates in proportion to $a(t)$, and therefore
as a function of $(t-t_0)/\tau_{sc}$ in a manner which 
depends on the box size. Thus the dependence on
box size in the evolution  of these systems can arise 
not just from contributions to the gravitational force 
due to the  periodic copies, but also through possible differences
in the dynamics due to the differences in force smoothing
in each case. Apart from such effects their
evolution should be identical when analysed in physical
units (of length and time). }

\begin{table}
\begin{center}
\begin{tabular}{|c|c|c|c|c|c|}
\hline
  Name  &  $\varepsilon$   & $a_{\rm end} $ \\
\hline 
   s5R0.3     &  1.1 $\times 10^{-3}$  & 100                       \\
   s5R0.2     &  7.4 $\times 10^{-4}$  & 67                       \\
   s5R0.1     &  3.7 $\times 10^{-4}$  & 33                      \\
    s5R0.05     &  1.65 $\times 10^{-4}$  & 16.8                       \\
     s5R0.025     &  8.25 $\times 10^{-5}$  & 8.6                       \\
 \hline
\end{tabular}
\end{center}
\caption{Parameter of simulations with $N=10^4$ particles for
various different initial system size $R$ (in units of the 
side of the periodic box). The ratio $\varepsilon/R$ is
the same as that in the simulation s5 in Table \ref{table-one}.
The parameter $a_{\rm end}$ is defined in the text.}
\label{table-three}
\end{table}

{In Fig.~\ref{fig_UP_DiffBoxes} is shown the evolution of the 
potential energy $U_p$, normalized to the modulus of its initial value, 
for the five simulations, as a function of the appropriately normalised
time determined by Eq.~(\ref{time-elapsed}), and  the evolution
of the virial ratio in the inset. In Fig.~\ref{fig_DP_DiffBoxes} 
is shown the density profile in each of the five simulations, at the time 
in each case corresponding to $a=10$ in the simulation s5 (and s5R0.1). 

Very strong dependence on the box size is manifest for the two largest
systems (of which the initial diameter is equal to, respectively,
$0.4$ and $0.6$ of the box size): both the energy of the virial
equilibrium attained, and the density profiles, are very significantly
different to those in the other cases. The smaller systems, on the
other hand, show a clear convergence for these quantities (and which
we have seen for the case s5R0.1 agree well with those obtained in the
open case).  The differences at late times in the deviations towards
higher values of $U_p$, associated (as can be seen in the inset) with
deviations from satisfaction of the virial condition, may clearly be
attributed to the differences in the force smoothing noted above: a
larger box size corresponds, at a given $(t-t_0)/\tau_{sc}$, to a
larger scale factor $a$, and therefore to a larger smoothing with
respect to the size of the system.  Indeed in the simulation s7 we saw
(Fig.~\ref{fig4} ) that significant deviation in the behaviour of
$U_p$ already at $a \sim 2-3$.  The simulations here have
$\epsilon/R_s$ initially exactly ten times smaller than in s7, and so
would be expected to show deviations in the range $a \sim 20-30$ as is
observed.

{Thus for the quantities we have focussed on above the effects due to
  the finite box size in the system of the initial size we have
  considered appear indeed to be very small.  We note, however, that
  considerable box size dependence is manifest even for the smaller
  boxes in other quantities. The amplitude of the oscillations during
  the initial relaxation to virial equilibrium are very significantly
  larger than in the smaller systems, for which the amplitude appears
  to converge approximately.  This suggests that the corresponding
  modes of oscillation of the structure about the virial equilibrium
  are enhanced very significantly by the gravitational coupling to the
  periodic copies.}

\begin{figure}
\vspace{1cm} { \par\centering
  \resizebox*{9cm}{8cm}
{\includegraphics*{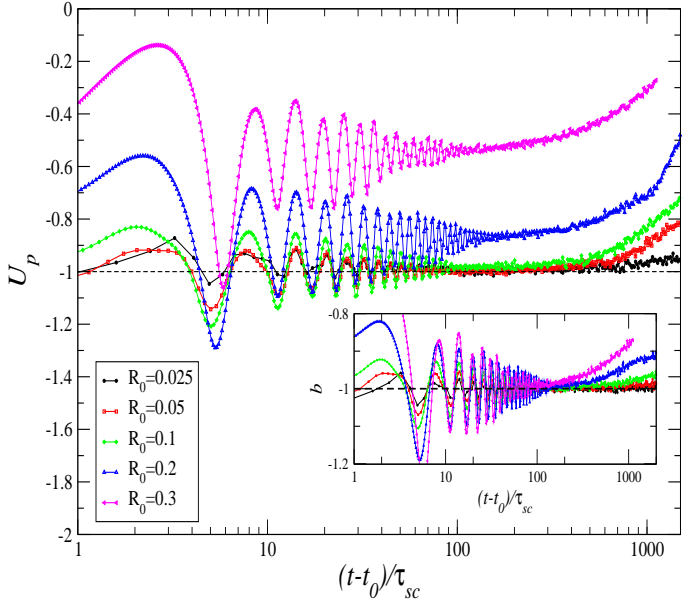}}
  \par\centering }
\caption{Evolution of the potential energy $U_p$ for the
  simulations in Table~\ref{table-three} with $N=10^{4}$,
  normalized to initial absolute value of $U_p$.}
\label{fig_UP_DiffBoxes}
\end{figure}

\begin{figure}
\vspace{1cm} { \par\centering
  \resizebox*{9cm}{8cm}
{\includegraphics*{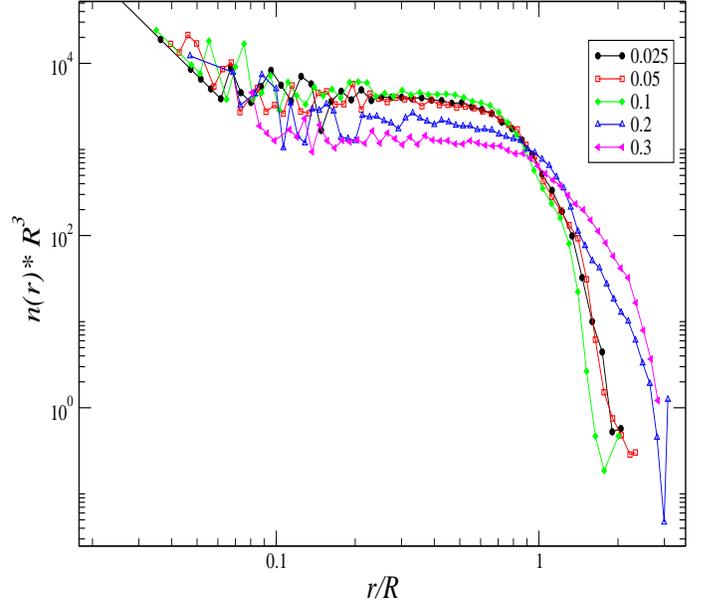}}
  \par\centering }
\caption{Density profiles for the
  simulations in Table~\ref{table-three} with $N=10^{4}$,
  at the normalized time corresponding to $a=10$ in s5R0.1}
\label{fig_DP_DiffBoxes}
\end{figure}

\section{Comparison with the Layzer-Irvine test}

{One possible test of the accuracy of a cosmological code is the
  so-called Layzer-Irvine (LI) test, derived from the equation of the
  same name which describes the variation of total energy in an
  expanding universe (see, e.g., \cite{peebles}).  Unlike energy
  conservation in non-expanding simulations, it is, however, a test
  which is rarely employed in practice by cosmological simulators as a
  control on their code, because it is not evident how to quantify the
  violation of the LI equation which may be tolerated\footnote{The
    GADGET2 user guide, for example, states that ``the cosmic energy
    integration is a differential equation, so a test of conservation
    of energy in an expanding cosmos is less straightforward that one
    may think, to the point that it is nearly useless for testing code
    accuracy.''}. Given that the test discussed in this paper is an
  independent one on the correctness of cosmological codes in a
  specific regime, it can in principle be used to ``calibrate" the LI
  test in this context. Likewise it is interesting to see whether it
  may be useful to employ both tests together.

In terms of the quantities defined above, the Layzer Irvine equation
may be written \citep{peebles}}
\begin{equation}
\frac{d}{dt}\left[a(K_p+U_p)\right] = - \dot{a} K_p \,.
\label{LIeqn}
\end{equation}
We thus define the quantity 
\begin{equation}
A(a) = \frac{a(K_p +U_p) + \int_1^a K_p da}{ K_p(1) +U_p (1)}
\label{LI-alpha}
\end{equation}
which should be equal to unity \footnote{The LI test remains
  applicable in this form for the infinite periodic system, if $U_p$
  is calculated with the corresponding two body potential. As this
  makes no significant difference to the results we give below, we
  continue to use $U_p$ as considered elsewhere in the paper
  (i.e. calculated without this modification to the two body
  potential).}.

In Fig.~\ref{fig_LItest} is shown the evolution of $A(a)$  for the 
simulations s1 to s7. We observe  immediately
that the two simulations in which $A(a)$ 
remains close to unity are precisely those, s5 and s6, 
which have been singled out by our test above as reproducing
best the required behaviour. On the contrary, all the other 
simulations which showed much greater deviation from stable
clustering also show larger deviations from unity of $A(a)$. 
More precisely, in all
cases where deviation of $A(a)$ from unity {\it by more than 
a few percent} is observed, the stable clustering test 
showed the results for the clustering in the system 
were completely incorrect and unphysical. 

The correlation between the information we deduced from
Fig.~\ref{fig4} and what we observe in Fig.~\ref{fig_LItest} is very
evident, and the reason for it very simple: the requirement that $U_p$
be constant is, {\it when the virial condition $b=-1$ is satisfied},
equivalent to the condition that both $K_p$, and therefore also the
sum $U_p + K_p$ corresponding to the energy in physical coordinates,
is constant.  In this case, as can be seen from Eqs.~(\ref{LIeqn}) and
(\ref{LI-alpha}), the LI test is also satisfied.  Thus it is clear
that the very strong deviations from stable clustering observed for
the simulations s1 to s4 stem from a poor integration of the equations
of motion, with very significant violation of energy conservation. As
discussed in Sect.~\ref{Parameters of expanding simulations}, such
numerical difficulties are expected to arise due to the precision
requirements of integrating accurately the hard two body collisions
present when $\varepsilon$ becomes very small. { We note that our
  results are very consistent with those of \cite{knebe_etal_2000} who
  have studied in detail the effects of such poorly integrated hard
  collisions: they lead to an artificial injection of energy into the
  system, increasing its size as some particles are sent into spurious
  higher energy orbits. We observe here indeed quite distinctly these
  effects both in the behaviour of the energy (which increases) and
  the profiles which stretch out further than they should (compared to
  stable clustering).}

For the very small $\varepsilon$ considerably greater precision than 
that employed (see Appendix \ref{appendix}) would be required 
to attain numerical convergence. We note that our results do suggest 
that, at given numerical precision, a lower bound on $\varepsilon$ may 
be expressed in a simple form like
Eqs.~(\ref{epsilon-lower-bound-1})-(\ref{epsilon-lower-bound-3}),
i.e., that the minimal $\varepsilon$ required scales linearly the size of
the structure $R_s$ : in Fig.~\ref{fig4} each of the simulations s1 to
s3, which have the same maximum time step, show an approximate plateau
in $U_p$, starting from a scale factor $a$ which increases roughly in
proportion to $1/ \varepsilon$.  Given that we observe in these
simulations that $R_s \sim 1/a$, this behaviour of $U_p$, 
indicating energy conservation, therefore sets in approximately 
at a fixed value of $\varepsilon/R_s$, in line with bounds of the form
of Eqs.~(\ref{epsilon-lower-bound-1})-(\ref{epsilon-lower-bound-3}).
A study of simulations with different $N$ would  be required to
establish which (if any) of the proposed scalings is the correct one
\footnote{{  \cite{knebe_etal_2000} propose bounds based on the
same simple argument given above for Eq.~(\ref{epsilon-lower-bound-1}).}}

We underline that the LI test for an expanding simulation is, just as
the test for the constancy of $U_p$, a weaker test than the test for
the stability of clustering: the latter tests for the collisionless
nature of the evolution, which is a different (and stronger)
requirement than energy conservation.  In practice, however, the
breakdown of energy conservation is often due to the difficulty of
integrating numerically with sufficient accuracy the collisional
dynamics (specifically, hard two body collisions), and therefore the
breakdown of the collisionless approximation is associated with the
violation of energy conservation. Such an association can always be
``undone" , in principle, by increasing sufficiently the accuracy of
the numerical integration. { In practice, however, it is very
  difficult to disentangle the two effects, and indeed studies up to
  now of two body collisionality in cosmological simulations
  (e.g. \cite{knebe_etal_2000, power_etal_2002, binney+knebe_2002})
  have not done so.}

In conclusion we find that the LI test is a very useful and relevant
one in the context of the present test of cosmological
simulations. Quite simply deviations of the dimensionless parameter
$A$ defined above by more than a few percent appear always to be
indicative of a grossly incorrect evolution. The crucial difference
with respect to its use for a full cosmological simulation, which, as
mentioned, has been found to be problematic, arises from the
difference in initial condition: for the very cold and almost
perfectly uniform initial conditions of cosmological simulations the
denominator in $A$ approaches zero, which makes it
difficult to calibrate the test.}

\begin{figure}
\vspace{1cm} { \par\centering
  \resizebox*{9cm}{8cm}
{\includegraphics*{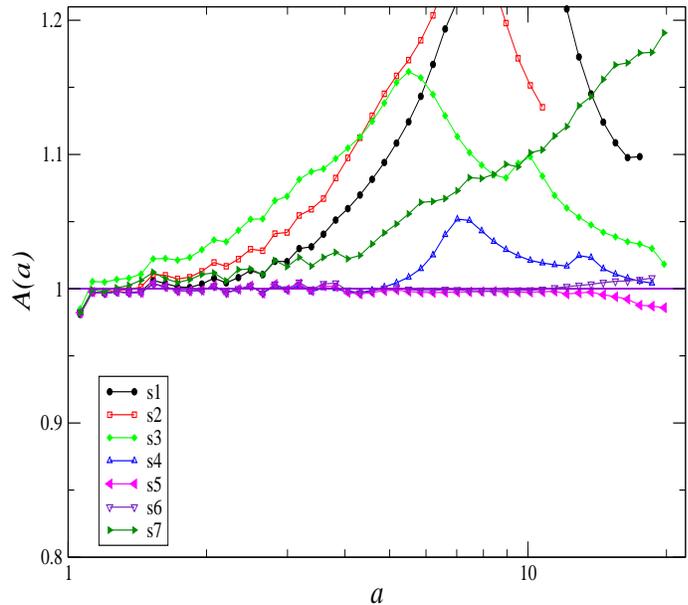}}
  \par\centering }
\caption{Layzer Irvine test for the simulations in Table~\ref{table-one}.
See text for the definition of $A(a)$. }
\label{fig_LItest}
\end{figure}

\section{Nature of discreteness effects and parametric scalings}
\label{Nature of discreteness effects and parametric scalings}

Our results above establish that the tests considered can clearly
detect and measure discreteness effects in $N$-body simulations, i.e.,
deviations of the results of such simulations from the desired
continuum limit.  We discuss now briefly what conclusions may be drawn
about the {\it nature} of these discreteness effects. More
specifically we discuss what we can conclude about the {\it parametric
  dependences} of these effects.

Discreteness effects here can be divided into two categories as
follows:
\begin{itemize}
\item``numerical discreteness effects" arising arise from limitations 
on precision in the integration of the $N$-body system with a given 
smoothed two-body potential, and
\item ``physical discreteness effects" arising from the 
use of a finite {\it particle density}, finite {\it force smoothing} 
and a finite {\it  periodic box}.
\end{itemize}

{The former are related to the choice of the numerical parameters
  controlling accuracy of the force calculation and time stepping in
  the code, while the latter are related to the number of particles
  $N$, the size of the force smoothing $\varepsilon$ and the size of
  the box $L$ \footnote{We do not consider here the starting red-shift
    $z_i$ for a simulation which is a parameter introduced in the
    $N$-body discretisation and on which discreteness effects may
    depend (see e.g. \cite{discreteness3_mjbm, knebe_etal_2009}).  As
    we study here the evolution only of non-linear structures from the
    time they form, we cannot constrain $z_i$ which can affect the
    evolution prior to this time.}. The two kinds of effects are {\it
    in practice} interrelated since the numerical precision required
  will depend typically on the values of the ``physical" discreteness
  parameters.  However, in order to understand the effects at play in
  cosmological simulation, it is useful to separate them in this
  way. One can then consider, on the one hand, the issue of numerical
  convergence at fixed values of $N$, $\varepsilon$ and $L$, and, on
  the other hand, the scalings with $N$, $\varepsilon$ and $L$, of the
  deviations from the desired physical behaviour, assuming ``perfect"
  numerical convergence.  It is the latter we consider here. Our
  results in Sect.~\ref{Results for different box sizes} showed up
  clearly the presence of effects related to the box size $L$, which,
  in line with expectations, decrease strongly as $L$ increases
  compared to the size of the simulated structure.  We do not pursue
  further tests here to establish exactly the associated scalings, but
  instead focus on the other two parameters.

{ Our results show clearly, as anticipated, that for larger values of
  the smoothing, deviation from the collisionless self-gravitating
  limit arises predominantly from the associated loss of spatial
  resolution. } This was most easily ``diagnosed" by the behaviour of
the virial ratio, which deviated clearly away from $b=-1$ towards less
negative values. As discussed in Sect.~\ref{Results for different box
  sizes} the behaviours observed are very consistent with the simple
bound Eq.~(\ref{epsilon-upper-bound}), with significant deviations
becoming easily visible (in potential energy and profiles) roughly
when $(\varepsilon/R_s) \sim 0.1$.  Using the fact that $R_s \propto
1/a$, we deduce that the scale factor at which we see the effects of
the finite resolution start to significantly modify the structure is
\begin{equation}
a_{res} \sim 10^{-1} \frac{R}{\varepsilon} 
\label{resolution-pred}
\end{equation}
where $R$ is the size of the structure at $a=1$
(and the result is valid for the case we consider
where $\varepsilon$ remains fixed in comoving coordinates).
 
The very large deviations from stable clustering we have observed in
our simulations with very small $\varepsilon$ are, as we have
discussed, apparently due to poor integration of hard two body
collisions. This problem is clearly diagnosed very well using analysis
of the $U_p$, or even more clearly using the LI test.  { As mentioned,
  such effects have been diagnosed and discussed in some detail
  notably in \cite{knebe_etal_2000}.}

{ In principle, as we have discussed, two body collisionality, {\it
    when integrated accurately}, can also contribute to the deviations
  from stable clustering we observe. Indeed in our simulations s5 and
  s6, which satisfy quite precisely the LI test, we see clear
  deviations which are similar qualitatively to those observed in the
  case of poorly integrated collisions: tails in the density profiles
  which become more extended in time and a hint of steepening of the
  inner density profile.  It is straightforward in our case to
  estimate the time scale for such two body effects using the well
  known results for the case of an open virialized system in a
  non-expanding space.} In this case numerical studies (see
e.g. \cite{Farouki+salpeter_1982, Farouki+salpeter_1994, theis_1998,
  theis+spurzem_1999, diemandetal_2body, gabriellietal_prl2010}) have
shown that the time scales for evolution of collisionless equilibria
is very consistent with the those estimated analytically (originally
by Chandrasekhar, see \cite{chandra43}) for two body collisions, given
by
\begin{equation}
\tau_{2body} \approx \kappa N \tau_{c}
\label{2body-relaxation-static}
\end{equation}
where $\tau_{c} $ is a characteristic crossing time for
the structure,  and $\kappa$ is a numerical factor incorporating 
the ``Coulomb logarithm"~\footnote{This logarithmic factor is simply
$\log (R/\varepsilon)$ when $\varepsilon$ is larger than the minimal 
impact factor required for the validity of the soft collision approximation.
This calculation is for the case of a {\it time  independent} smoothing.}.
Modulo box size effects, which we have seen are small, the only
difference between these open systems and the one we are studying
should arise from the difference in the smoothing, which in
our cosmological code simulations is fixed in comoving coordinates 
and therefore varies in time (increases) in
physical coordinates.  The two body relaxation time is, however, only
logarithmically sensitive to the lower cut-off in the two body
interaction, and temporal variation of the smoothing $\varepsilon$
will lead therefore to modification of the numerical prefactor $\kappa$ 
in the calculation, with at most very weak dependence on $N$.
Using Eq.~(\ref{time-elapsed}) with Eq.~(\ref{2body-relaxation-static}) we
thus estimate that two body relaxation will start to cause deviations in 
our numerically well converged test simulations (i.e. s5 and s6 above) 
when 
\begin{equation}
a_{2body} \sim \left[\frac{\kappa (\varepsilon) }{\sqrt{\delta} }
  \,N\right]^{2/3}
\label{relaxation-pred}
\end{equation}
where we have taken $\tau_{c} \sim \tau_{sc}$.

Values of $\kappa$ of order those measured in open simulations (with
smoothing fixed in physical coordinates), typically a little smaller
than unity, thus give an estimate for $a_{2body}$ quite consistent
with the hypothesis that two body collisionality should account for
the observed deviations from stable clustering in the simulations in
which the energy is conserved well. { We note that these results also
  appear to be very consistent with studies such as
  \cite{power_etal_2002, binney+knebe_2002} which have detected such
  effects in cosmological simulations. } Clearly however a full study
of the $N$-dependence of the results of our test, which we do not
undertake here, would be required to establish whether the scaling
Eq.~(\ref{relaxation-pred}) is indeed observed.  It would be
instructive to couple such a study also to one including analysis with
other indicators of collisionality (e.g. measures of diffusion in
velocity space like those employed in \cite{diemandetal_2body}, { or
  using two mass species as in \cite{binney+knebe_2002}}).

An important practical question in numerical simulation is whether
there is an {\it optimal} value of force smoothing in cosmological N
body simulations. The question may be posed either with respect to
some set of numerical constraints (quantified e.g. in terms of bounds
on the numerical parameters controlling accuracy of integration), or
in abstraction from such limitations.  The relations
Eq.~(\ref{relaxation-pred}) and Eq.~(\ref{resolution-pred}) can be
clearly combined, in principle, to provide a prescription for an
optimal value of the latter kind: although we have not determined
explicitly, it is clear that $\kappa (\varepsilon)$ is a monotonically
(a priori logarithmically) increasing function of $\varepsilon$, while
the coefficient in Eq.~(\ref{resolution-pred}) is monotonically
decreasing. Thus by varying $\varepsilon$ at fixed $N$ an optimal
value can be found which maximizes the scale factor at which
discreteness effects modify the evolution.  A much more extensive
study, notably of $N$ dependence, would clearly be necessary to
determine the scaling with $N$ of such an optimum.  It would be
interesting to determine, in particular, whether it turns out to be
that derived by \cite{merritt_1996, dehnen_2001} from simple
considerations of the optimization of the representation of the
continuum (mean-field, Newtonian) forces.  It would evidently be
interesting also to compare with optimization criteria derived in
various numerical convergence studies in the literature (e.g.
\cite{diemandetal_2body, power_etal_2002}).}

\section{Practical guide to use of the test}
\label{Quantitative constraints for cosmological simulations}

The numerical study we have reported establishes that the
test considered can provide non-trivial constraints on the accuracy 
of cosmological simulations.  To facilitate the exploitation of the test 
in practice by cosmological simulators we now summarize in a recipe 
form how it could be implemented. The results of such tests will,
as we discuss,  inevitably depend  strongly on the details of the 
particular simulations (cosmological model, time and length scales 
probed, quantities of interest and desired level of precision etc).
We therefore do not attempt to summarize the information
which can be obtained from the test in some simple set of rules.
Instead we propose to simulators an instrument they can use to 
assess themselves the reliability of the results of their codes.

{Let us suppose a simulator intends to run a (dissipationless)
  cosmological simulation, which has $N_{\rm 0}$ particles of mass $m$
  in a box of side $L_{\rm 0}$.  An initial perturbation spectrum
  $P(k)$ is given at the starting red-shift $z_i$, and a cosmological
  model specifying the scale-factor (e.g. $\Lambda$ CDM).  Based on
  criteria at his disposal, he chooses a set of numerical parameters
  (time step and force accuracy criteria, smoothing
  $\varepsilon$). The test we have considered for the accuracy of
  simulation of halos containing $N$ particles (and of mass $M=mN$)
  can be generalized as follows:
\begin{itemize}

\item[1.]  Consider a simulation identical for all
numerical parameters to the full simulation, {\it except
for the box size}, which is rescaled so that
\begin{equation}
L=L_0 \left( \frac{N}{N_0} \right) ^{1/3} 
\label{box-size}
\end{equation}
This condition means that the mean matter density of the universe is
equal to that in the full simulation, { and therefore, in particular,
  the ratio $\varepsilon/\ell$ has the same value as in the full
  simulation.}

\item[2.] Distribute the $N$ particles in a spherical
region of radius $R= L (3\rho_0/4\pi \rho_v)^{1/3}$ 
where $\rho_v$ is the initial density of the halo. 
For $\rho_v \approx 200 \rho_0$ this corresponds
to $R \approx 0.1$ (as in our simulations s1-s7 above).
 
\item[3.]  Assign velocities with some simple velocity distribution
(e.g. uniform or gaussian) and normalize them so that the initial
virial ratio is unity.  
 
\item[4.] Run the cosmological code (with the chosen 
numerical parameters) starting from $z_v (M)$, 
the (estimated) maximum redshift for virialization of a halo 
of mass $M$  in the model studied. 

\item[5.] Check that the  potential energy 
$U_p$   and virial ratio $b$ display the expected 
physical behaviour (low amplitude decaying oscillations 
about a constant value). Systematic deviation of 
the virial ratio from $b=-1$ towards less negative
values indicates that the upper resolution on
smoothing is violated. Plot also the parameter $A(a)$ 
for the LI test. If deviations from unity are at the level of
more than a few percent, the parameter choices are 
inappropriate to simulate over the corresponding 
time-scale.

\item[6.]  To obtain more precise limits on the precision of 
halo profiles (in the cases where the previous tests are
reasonably well satisfied) apply first the test for stability
of clustering. Further quantitative limits can be obtained
by comparison with the equilibrium profile obtained
from a high-resolution simulation of the same 
initial conditions in open boundary conditions.

\end{itemize}

Our simulations s1-s7 reported above test the case $N=10^4$ for a
range of values of $\varepsilon$ (and other numerical parameters) up
to red-shift of about twenty. Following the prescription above we
would exclude all but s5 and s6 using the analysis of the energies and
LI test beyond a red-shift of a few (cf. Fig.~\ref{fig4} and
Fig.~\ref{fig_LItest}).  The conclusion of our further analysis (step
6) above was that significant departures from stable clustering (and
from the equilibrium template) were observable, increasing as a
function of redshift. Beyond a redshift of order ten, we could
conclude, for example, that $50\%$ precision on profiles of halos over
two decades in scale is not attainable with $N=10^4$.  To determine
how many particles (and what numerical parameters) would lead to such
a level of precision could be determined by performing the test for
larger $N$.

Several variants of the test as described above could provide further 
or more precise constraints: 
\begin{itemize}

\item Different initial conditions could be used in steps 2 and 3. In
  particular rather than a structure close to, but not at, virial
  equilibrium, one could take instead a structure which is already at
  equilibrium. Such an initial condition could be prepared
  numerically, or set up directly using the Eddington formula (see,
  e.g., \cite{muldrew_etal_2011, Kazantzidis_2004}).  It would be
  interesting notably to study equilibrium halos with profiles
  like those typically measured in simulations (e.g. NFW halos).
  Alternatively initial conditions could be obtained directly from the
  full cosmological simulation, by extracting typical halos at about
  the time they virialize. Techniques to do this have been developed
  in the context of convergence studies in which individual halos are
  identified and resimulated at higher resolution(see,
  e.g. \cite{power_etal_2002} and references therein).

\item To separate out possible effects arising from the box size, the
  test could be repeated in boxes of different sizes, and specifically
  in a box of side $L_0$. These are important in particular when
  comparison with a simulation in open boundary conditions is made.
  As the change in the box size leads to a change in the mean density
  (by a factor of $N/N_0$ for the case of a box of size $L_0$), this
  must be accounted for in applying the test. As discussed in
  Sect.~\ref{Results for different box sizes}, in an EdS cosmology
  this can done, modulo effects due to the force smoothing, by a
  simple rescaling of the time. If the role of smoothing is the focus
  of the test, and/or the cosmology employed is non-scale free
  (e.g. $\Lambda$CDM,) the code would need to be modified ``by hand"
  to impose the evolution of the scale factor corresponding to the
  cosmological simulation.

 \item Another variant of the test would be, using further the
   framework of the spherical collapse model for halo formation, to
   start from the time of turn-around. At this time the spherical
   overdensity is already sufficiently large ($\sim 5-6$) that one
   should be able to consider the system to a good approximation as
   isolated. The initial condition would thus be taken as a uniform or
   quasi-uniform sphere with physical velocities equal to zero.  Given
   that the scale $R$ in step 1 would be comparable to the box size,
   it would be appropriate also to study simulations in a larger box
   as discussed in the previous paragraph.

\end{itemize}
}

\section{Conclusions and discussion}

The central point of this paper is the introduction of a simple test on
the reliability of cosmological $N$-body simulations in the non-linear regime.
We have shown with a very simple implementation of the test that it
can constrain strongly the choice of the unphysical parameters introduced 
by the $N$-body method, {  given desired precision criteria on the
properties of virialized structures. More specifically we have illustrated 
that the test can  determine a window for an appropriate force 
smoothing $\varepsilon$ for simulations with a given time stepping 
accuracy, and any given $N$ in a virialized structure.  At larger
$\varepsilon$ the loss of resolution was clearly seen to be
the dominant effect, while at small $\varepsilon$ the effects of
two body collisionality become the primary cause of deviation.
{\it In contrast to most other methods explored previously in the 
literature these effects are detected and quantified by comparing 
the evolution of the test simulation with an exact behaviour,
rather than by a comparison between different codes.}
  
We have not attempted here, and indeed it is not our goal, to try to
use the test to derive some set of simple short-hand rules that could
be used to, say, choose $\varepsilon$ in a given cosmological
simulation (or compared with other prescriptions which have been given
in the literature). Rather it is the simulator who should apply the
test to derive what his constraints are, with respect to his own
precision requirements.  Nevertheless it is interesting to comment a
little more on what our specific test, of simulations with fixed
comoving softening, indicates. Smoothing in simulations with fixed
comoving softening are characterised by the sole parameter
$\varepsilon/\ell$.  We can compare the smoothings we have considered
(Table~\ref{table-one}) to typical values employed in large volume
cosmological simulations: the smoothing in s5, s6 and s7 correspond,
respectively, to $\varepsilon/\ell \approx 1/125, 1/100, 1/12$, while
e.g., \cite{springel_05} uses $\varepsilon/\ell \approx 1/50$, and
\cite{smith} $\varepsilon/\ell \approx 1/16$.  From the results
discussed above it is clear that application of the test for these
specific values can give very strong indications of the reliability
and/or precision of the properties of halos in such
simulations. Indeed we note that, using Eq.~(\ref{ratio-densities}),
our approximate derived constraint Eq.~(\ref{resolution-pred}) for the
red-shift range over which a structure of $N$ particles may be
resolved, can be rewritten as
\begin{equation}
a_{res} \sim 10^{-2} \frac{\ell}{\varepsilon} \, N^{1/3} \,. 
\label{resolution-pred-cosmo}
\end{equation}
Thus, for example, at $\frac{\ell}{\varepsilon} \sim 10^{-2}$
resolution constraints become relevant for all
objects of less than $N \sim 10^3$ formed at 
or before a redshift of ten.
 
Comparing this with Eq.~(\ref{relaxation-pred}),
in which the factor $\kappa (\varepsilon)$ has a priori very 
weak (logarithmic) dependence on $N$,  leads to the 
conclusion that,  for given $\varepsilon/\ell$, two body 
collisionality will be the dominant discreteness effect at 
small $N$, while  for larger $N$ it is the resolution limit 
associated with $\varepsilon$ which is the relevant one. 
Thus, for an $\varepsilon/\ell$ which remains fixed 
in comoving coordinates, there is an inevitable trade-off 
between resolution and the introduction of spurious 
discreteness effects.  We have considered solely simulations 
with fixed comoving softening, but the test can be applied 
without modification (as described in detail in that last section) 
to any other kind of code. Indeed it could provide 
very useful constraints on optimal smoothing strategies
in codes with an adaptative smoothing, which indeed
aim to optimize spatial resolution while keeping two 
body collisionality under control (see e.g. \cite{knebe_etal_2000}).}

We finally remark that while the test discussed here can, as we have
shown, provide constraints on the parameters which must be respected
in order to reproduce the desired continuum limit, it does not show
that satisfaction of these constraints {\it guarantee} the same
property, i.e., the test provides necessary, but not sufficient,
conditions to guarantee the correctness of the numerical results.  {
  Firstly the test only constrains the choice of parameters in the
  strongly non-linear regime. This means, notably, that it cannot say
  anything about the appropriateness of the use of an $\varepsilon <
  \ell$, which has been shown to introduce discreteness effects in the
  early time evolution, and which may or may not distort the
  subsequent evolution (\cite{splinter_1998, discreteness3_mjbm,
    romeo08, knebe_etal_2000}).  Nor can it, as noted, provide
  constraints on the choice of the starting redshift of simulations
  (\cite{discreteness3_mjbm, knebe_etal_2009}).  Further, even in the
  non-linear regime, } it is clear that considerable caution should be
adapted in supposing the constraints derived from this test guarantee
a faithful representation of the collisionless evolution: it shows
that halos with {\it less} than some number of particles will
necessarily suffer from effects of discreteness which modify strongly
their density profiles: as we have seen in our study, {\it the use of
  a smoothing which is slightly too small leads at the end of the
  simulation to a completely incorrect profile}.  Further the
characteristic physical scale of this modification is the size of the
structure, which has no direct relation to the smoothing scale
$\varepsilon$. In the case of an inappropriately large $\varepsilon$
(simulation s7 above) we saw that a halo can even be very considerably
larger than it should be.  How the presence of such spurious
clustering modifies the evolution of the whole system is very
unclear. Indeed given that clustering in currently favored
cosmological models is hierarchical in nature, the possibility that
such error may feed through different scales is a major concern.

We thank Fran\c{c}ois Sicard for useful discussions, and the two
anonymous referees for many invaluable criticisms and suggestions.

\appendix
\section{Further details on simulations}
\label{appendix}

Besides the choice of $\varepsilon$ on which we have focussed above,
simulation with GADGET requires one to fix several other numerical
parameters. In our choices we have followed the guidelines given by
the GADGET user's guide, and also performed several tests indicate
that our results appear to be reasonably stable with respect to these
choices. Specifically we have considered:

\begin{itemize}
\item {\tt ErrTolIntAccuracy}, a dimensionless parameter which
  controls the accuracy of the time-step criterion.  The value
  suggested by the GADGET user's guide is 0.025, and we have done
  numerical tests in various cases reducing it by a factor ten without
  any detectable difference in the results reported here.

\item {\tt MaxSizeTimestep}, which specifies the maximum allowed
  time-step for cosmological simulations as a fraction of the current
  Hubble time. According to the GADGET user's guide a value of 0.025
  is usually accurate enough for most cosmological runs. We have
  found our results to be stable in tests using considerably smaller
  values, down to as small as $10^{-5}$ in some cases.
  
\item
{\tt ErrTolTheta} is the accuracy criterion (the opening angle
$\theta$) of the tree algorithm if the standard Barnes \& Hut (BH)
opening criterion is used. The suggested value is 0.7 and we have
considered values down to 0.1, again finding stable results in
the tested cases.

\end{itemize} 

We have also performed some tests comparing different realizations
of the initial conditions in various cases, again with stable results.

{We emphasize that despite these numerous tests, we have come to the
  conclusion through our analysis using the test studied in the paper
  that the four simulations with smaller $\varepsilon$ are in fact not
  numerically converged, and are characterized by very significant
  violations of energy conservation due to poorly integrated two body
  collisions.  Thus sufficient further extrapolation of the numerical
  parameters beyond what we have considered must lead to very
  different results in these cases.}

For completeness we give in Table~\ref{tab1} I the values of the
numerical parameters used in the simulations reported in the body of
the article. {For the non-expanding simulation in open boundary
  conditions we have used (see also \cite{sl2012} for further
  details), a force softening $\varepsilon= 0.007 R$. This means that
  $\varepsilon$ is very much smaller at all times than the size of the
  structure. In addition we have chosen $\eta=0.01$, and have used the
  new GADGET cell opening criterion with a high force accuracy of
  $\alpha_F = 0.001$. Energy is conserved to within less than
  $10^{-3}$ up to $12 \tau_{sc}$.  In \cite{Joyce_etal_MNRAS2009}
  extensive tests of the dependence on $\varepsilon$ were performed,
  for the much more constraining case of evolution from the same
  initial condition but with initial velocities set to zero.  These
  tests lead to the conclusion that no sensible dependence is observed
  up to this time (notably in the density profile) unless the ratio
  $\varepsilon/R_s$ becomes large.}

\begin{table}
\begin{center}
\begin{tabular}{|c|c|c|c|c|c|}
\hline
  Simulation & ETIA    & MaxTimeStep & ErrTolTheta      \\
\hline 
   s1                & 0.025  & 0.01        & 0.7              \\
   s2                & 0.0025 & 0.01        & 0.7              \\
   s3                & 0.0025 & 0.01        & 0.7              \\
   s4                & 0.025  & 0.0001      & 0.7              \\
   s5                & 0.025  & 0.0001      & 0.7              \\
   s6                & 0.025  & 0.00001     & 0.1              \\
   s7                & 0.025  & 0.0001      & 0.7              \\
\hline 
  s1N5e4             & 0.025  & 0.01        & 0.7              \\
  s1N1e5             & 0.025  & 0.01        & 0.7              \\
\hline 
  s5N1e3             & 0.025  & 0.0001      & 0.7              \\
  s6N1e3             & 0.025  & 0.00001     & 0.1              \\  
\hline
\end{tabular}
\end{center}
\caption{Values of the three numerical parameters (ErrTolIntAccuracy,
  MaxTimeStep and ErrTolTheta) adopted in each of the simulations
  reported in the text.}
\label{tab1}
\end{table}



\begin{thebibliography}{}

\bibitem[\protect\citeauthoryear{Bagla}{Bagla}{2005}]{bagla_review}
Bagla J.,  2005, Curr. Sci., 88, 10883

\bibitem[\protect\citeauthoryear{{Binney} \& {Knebe}}{{Binney} \&
  {Knebe}}{2002}]{binney+knebe_2002}
{Binney} J.,  {Knebe} A.,  2002, Mon. Not. R. Astron. Soc., 333, 378

\bibitem[\protect\citeauthoryear{Chandrasekhar}{Chandrasekhar}{1943}]{chandra4%
3}
Chandrasekhar S.,  1943, Rev. Mod. Phys., 15, 1

\bibitem[\protect\citeauthoryear{{Dehnen}}{{Dehnen}}{2001}]{dehnen_2001}
{Dehnen} W.,  2001, Mon. Not. R. Astron. Soc., 324, 273

\bibitem[\protect\citeauthoryear{Diemand, Moore, Stadel \& Kazantzidis}{Diemand
  et~al.}{2004}]{diemandetal_2body}
Diemand J.,  Moore B.,  Stadel J.,    Kazantzidis S.,  2004, Mon. Not. R.
  Astron. Soc., p.~977

\bibitem[\protect\citeauthoryear{{Efstathiou}, {Frenk}, {White} \&
  {Davis}}{{Efstathiou} et~al.}{1988}]{efstathiou_88}
{Efstathiou} G.,  {Frenk} C.~S.,  {White} S.~D.~M.,    {Davis} M.,  1988, Mon.
  Not. R. Astron. Soc., 235, 715

\bibitem[\protect\citeauthoryear{{Farouki} \& {Salpeter}}{{Farouki} \&
  {Salpeter}}{1982}]{Farouki+salpeter_1982}
{Farouki} R.~T.,  {Salpeter} E.~E.,  1982, Astrophys. J., 253, 512

\bibitem[\protect\citeauthoryear{{Farouki} \& {Salpeter}}{{Farouki} \&
  {Salpeter}}{1994}]{Farouki+salpeter_1994}
{Farouki} R.~T.,  {Salpeter} E.~E.,  1994, Astrophys. J. Lett, 427, 676

\bibitem[\protect\citeauthoryear{Gabrielli, Baertschiger, Joyce, Marcos \&
  Sylos~Labini}{Gabrielli et~al.}{2006}]{gabrielli_06}
Gabrielli A.,  Baertschiger T.,  Joyce M.,  Marcos B.,    Sylos~Labini F.,
  2006, Phys. Rev., E74, 021110

\bibitem[\protect\citeauthoryear{{Gabrielli}, {Joyce} \& {Marcos}}{{Gabrielli}
  et~al.}{2010}]{gabriellietal_prl2010}
{Gabrielli} A.,  {Joyce} M.,    {Marcos} B.,  2010, Phys. Rev. Lett., 105,
  210602

\bibitem[\protect\citeauthoryear{Gabrielli, Sylos~Labini, Joyce \&
  Pietronero}{Gabrielli et~al.}{2004}]{book}
Gabrielli A.,  Sylos~Labini F.,  Joyce M.,    Pietronero L.,  2004, Statistical
  Physics for Cosmic Structures.
Springer

\bibitem[\protect\citeauthoryear{{Heggie} \& {Hut}}{{Heggie} \&
  {Hut}}{2003}]{heggie+hut_book}
{Heggie} D.,  {Hut} P.,  2003, {The Gravitational Million-Body Problem: A
  Multidisciplinary Approach to Star Cluster Dynamics}.
Cambridge

\bibitem[\protect\citeauthoryear{{Heitmann}, {Ricker}, {Warren} \&
  {Habib}}{{Heitmann} et~al.}{2005}]{heitmann_etal_2005}
{Heitmann} K.,  {Ricker} P.~M.,  {Warren} M.~S.,    {Habib} S.,  2005,
  Astrophy. J. Supp, 160, 28

\bibitem[\protect\citeauthoryear{Joyce, Marcos \& Baertschiger}{Joyce
  et~al.}{2008}]{discreteness3_mjbm}
Joyce M.,  Marcos B.,    Baertschiger T.,  2008, Mon. Not. R. Astron. Soc.

\bibitem[\protect\citeauthoryear{{Joyce}, {Marcos} \& {Sylos Labini}}{{Joyce}
  et~al.}{2009}]{Joyce_etal_MNRAS2009}
{Joyce} M.,  {Marcos} B.,    {Sylos Labini} F.,  2009, Mon. Not. R. Astron.
  Soc., 397, 775

\bibitem[\protect\citeauthoryear{{Kazantzidis}, {Magorrian} \&
  {Moore}}{{Kazantzidis} et~al.}{2004}]{Kazantzidis_2004}
{Kazantzidis} S.,  {Magorrian} J.,    {Moore} B.,  2004, Astrophys. J., 601, 37

\bibitem[\protect\citeauthoryear{Knebe, Kravtsov, Gottl\"ober \& Klypin}{Knebe
  et~al.}{2000}]{knebe_etal_2000}
Knebe A.,  Kravtsov A.,  Gottl\"ober S.,    Klypin A.,  2000, Mon. Not. Roy.
  Astron. Soc., 317, 630

\bibitem[\protect\citeauthoryear{{Knebe}, {Wagner}, {Knollmann}, {Diekershoff}
  \& {Krause}}{{Knebe} et~al.}{2009}]{knebe_etal_2009}
{Knebe} A.,  {Wagner} C.,  {Knollmann} S.,  {Diekershoff} T.,    {Krause} F.,
  2009, Astrophys. J., 698, 266

\bibitem[\protect\citeauthoryear{{Kravtsov}, {Klypin} \& {Khokhlov}}{{Kravtsov}
  et~al.}{1997}]{kravtsov_1997}
{Kravtsov} A.~V.,  {Klypin} A.~A.,    {Khokhlov} A.~M.,  1997, Astrophys. J.
  Supp., 111, 73

\bibitem[\protect\citeauthoryear{{Merritt}}{{Merritt}}{1996}]{merritt_1996}
{Merritt} D.,  1996, Astron. Jour., 111, 2462

\bibitem[\protect\citeauthoryear{{Muldrew}, {Pearce} \& {Power}}{{Muldrew}
  et~al.}{2011}]{muldrew_etal_2011}
{Muldrew} S.~I.,  {Pearce} F.~R.,    {Power} C.,  2011, Mon. Not. R. Astron.
  Soc., 410, 2617

\bibitem[\protect\citeauthoryear{Peebles}{Peebles}{1980}]{peebles}
Peebles P. J.~E.,  1980, {The Large-Scale Structure of the Universe}.
Princeton University Press

\bibitem[\protect\citeauthoryear{{Power}, {Navarro}, {Jenkins}, {Frenk},
  {White}, {Springel}, {Stadel} \& {Quinn}}{{Power}
  et~al.}{2003}]{power_etal_2002}
{Power} C.,  {Navarro} J.~F.,  {Jenkins} A.,  {Frenk} C.~S.,  {White} S.~D.~M.,
   {Springel} V.,  {Stadel} J.,    {Quinn} T.,  2003, Mon. Not. R. Astron.
  Soc., 338, 14

\bibitem[\protect\citeauthoryear{{Romeo}, {Agertz}, {Moore} \&
  {Stadel}}{{Romeo} et~al.}{2008}]{romeo08}
{Romeo} A.~B.,  {Agertz} O.,  {Moore} B.,    {Stadel} J.,  2008, Astrophys. J.,
  686, 1

\bibitem[\protect\citeauthoryear{Smith, Peacock, Jenkins, White, Frenk, Pearce,
  Thomas, Efstathiou \& Couchman}{Smith et~al.}{2003}]{smith}
Smith R.~E.,  Peacock J.~A.,  Jenkins A.,  White S. D.~M.,  Frenk C.~S.,
  Pearce F.~R.,  Thomas P.~A.,  Efstathiou G.,    Couchman H. M.~P.,  2003,
  Mon. Not. R. Astron. Soc., 341, 1311

\bibitem[\protect\citeauthoryear{Splinter, Melott, Shandarin \& Suto}{Splinter
  et~al.}{1998}]{splinter_1998}
Splinter R.~J.,  Melott A.~L.,  Shandarin S.~F.,    Suto Y.,  1998, Astrophys.
  J., 497, 38

\bibitem[\protect\citeauthoryear{Springel et~al.,}{Springel
  et~al.}{2005}]{springel_05}
Springel V.,  et~al., 2005, Nature, 435, 629

\bibitem[\protect\citeauthoryear{{Springel}, {Yoshida} \& {White}}{{Springel}
  et~al.}{2001}]{springel_etal_2001}
{Springel} V.,  {Yoshida} N.,    {White} S.~D.~M.,  2001, New. Astron., 6, 79

\bibitem[\protect\citeauthoryear{{Sylos Labini}}{{Sylos Labini}}{2012}]{sl2012}
{Sylos Labini} F.,  2012, Mon. Not. R. Astron. Soc., 423, 1610


\bibitem[\protect\citeauthoryear{{Teyssier}}{{Teyssier}}{2002}]{teyssier_2002}
{Teyssier} R.,  2002, Astron. Astrophys., 385, 337

\bibitem[\protect\citeauthoryear{{Theis}}{{Theis}}{1998}]{theis_1998}
{Theis} C.,  1998, Astron. Astrophys., 330, 1180

\bibitem[\protect\citeauthoryear{Theis \& Spurzem}{Theis \&
  Spurzem}{1999}]{theis+spurzem_1999}
Theis C.,  Spurzem R.,  1999, Astron. Astrophys., 341, 361

\bibitem[\protect\citeauthoryear{Worrakitpoonpon}{Worrakitpoonpon}{2011}]{tira%
wut-thesis}
Worrakitpoonpon T., , 2011, PhD. Thesis, Universit\'e Pierre et Marie Curie

\end{thebibliography}

\end{document}